
\def\etal{{\it et al.}\hskip 2.5pt}
\def\cf{{\it cf.}\hskip 2.5pt}
\font\machm=cmsy10 \skewchar\machm='60

\def\refset{\parindent=0pt\hangafter=1\hangindent=1em}
\magnification=1200
\hsize=5.80truein
\hoffset=1.20truecm
\newcount\eqtno
\eqtno = 1
\parskip 3pt plus 1pt minus .5pt
\baselineskip 21.5pt plus .1pt
%
%
%
%
%
\centerline{ \  }
\vskip 0.35in
\centerline{\bf BACKGROUND X-RAY EMISSION FROM HOT GAS}
\centerline{\bf IN CDM AND CDM$+\Lambda$ UNIVERSES: SPECTRAL SIGNATURES}
\vskip 0.4in
\centerline{Renyue Cen$^1$, Hyesung Kang$^{1,2}$, Jeremiah P. Ostriker$^1$ and
Dongsu Ryu$^{1,3}$}
\vskip 0.3in
\centerline{email: cen@astro.princeton.edu}
\vskip 0.2in
 \centerline{In press {\it The Astrophysical Journal}}
\vskip 1cm
\line{$^1$ Princeton University Observatory, Princeton, New Jersey 08544
\hfill}
\line{$^2$ Department of Earth Sciences, \hfill}
\line{\qquad         Pusan National University, Pusan, Korea \hfill}
\line{$^3$ Department of Astronomy and Space Science,
\hfill}
\line{\qquad         Chungnam National University, Daejun, Korea\hfill}
\vfill\eject

\centerline{\bf ABSTRACT}

We present a new treatment of two popular models for the
growth of structure, examining the X-ray emission from hot
gas with allowance for spectral line emission from various atomic
species, primarily ``metals".
The X-ray emission from the bright cluster sources
is not significantly changed from prior work and,
as noted earlier,
shows the CDM$+\Lambda$ model (LCDM) to be
consistent but the standard, COBE normalized model (SCDM)
to be inconsistent with existing observations --- after
allowance for still the considerable numerical modelling uncertainties.

But we find one important new result.
Radiation in the softer band  0.5-1.0keV
is predominantly emitted by gas far from cluster centers
(hence ``background").
This background emission dominates over the cluster
emission below 1keV and
observations of it should show clear spectral signatures
indicating its origin.
In particular  the ``iron blend"
should be seen prominantly in this spectral bin
from cosmic background hot gas
at high galactic latitudes and should show
shadowing against the SMC indicating its extragalactic origin.
Certain OVII lines also provide a signature of this gas which
emits a spectrum characteristic of $10^{6.6\pm 0.6}$K
gas.
Recent ASCA observations of the X-ray background tentatively
indicate
the presence of component
with exactly the spectral features we predict here.

\noindent
Cosmology: large-scale structure of Universe
-- hydrodynamics
-- Radiation Mechanisms: Bremsstrahlung
-- Radiation Mechanisms: lines
-- X-ray: general
\vfill\eject

\centerline{1. INTRODUCTION}

In the three preceding papers, Kang \etal (1994), Bryan \etal (1994),
Cen \& Ostriker (1994 ``LCDM"),
we investigated the X-ray cluster properties in
the standard COBE-normalized cold dark matter model
and a CDM model with a cosmological constant,
under the assumption that bremsstrahlung was the
primary emission process.
While this assumption is an excellent approximation
for hard X-ray bands ($h\nu > 2$keV),
which characterize the emission from the great clusters,
its validity in softer bands has to be examined more carefully.
In this paper, we include detailed X-ray line emissions
from twelve major elements
(He, C, N, O, Ne, Mg, Si, S, Ar, Ca, Fe, Ni),
in addition to bremssrahlung due to these elements,
and we shift focus to the lower luminosity,
lower temperature clusters, filaments and sheets
that would normally be considered a part of the
``X-ray background".
We compare the X-ray cluster
luminosity functions and X-ray background
radiation fields
for two cases: with and without line emission
at different bands, for both cosmological models.
The issue of cooling flows in cluster centers is also addressed
in an approximate fashion,
and detailed X-ray spectra of a few bright X-ray clusters are presented.
Finally, we compute the ``background" emission
from hot gas in the two models --
the line and continuum emission
from gas far from the centers of rich clusters.
Our primary new result is that
the extragalactic background radiation field due to
hot gas far from the centers of rich clusters
is surprisingly large and it
dominates over other (diffuse) sources at energies below 1keV.

Very recent observations (Gendreau \etal 1994)
of the spectrum of the X-ray background
clearly indicate a thermal component from evidence
for OVII and OVIII lines.
The data also show ``an excess above the extrapolation of
the single power law model below 1keV"
of exactly the nature we describe in section 3.2.

In this paper we concentrate on the spectral signature of the background gas.
A subsequent paper (Cen \etal 1995)
will focus on the spatial distribution;
we note here in anticipation of that paper that
the angular autocorrelation function of the hot gas in our
simulations does (due to the background gas component)
show the extended features (up to several degrees) seen
in the real data (Soltan \etal 1995).

This present paper is organized as follows:
section 2 describes our method
of including line emission processes,
\S 3 gives the results and comparisons with observations,
and our conclusions are assembled in \S 4.

\medskip
\centerline{2. METHOD}
\medskip
The cosmological models which we are investigating here
are the same ones that we studied previously:
1) the standard, COBE normalized cold dark matter model (SCDM) with $\Omega=1$,
$H=50$km/s/Mpc, $\sigma_8=1.05$, $\Omega_b=0.06$ (Kang \etal 1994);
2) a COBE normalized, flat cold dark matter model with a cosmological
constant (LCDM) with $\Omega=0.45$,
$\lambda=0.55$, $H=60$km/s/Mpc, $\sigma_8=0.77$, $\Omega_b=0.043$
(Cen \& Ostriker 1994).
Both model simulations are performed in a box of size
$85h^{-1}$Mpc with $270^3$ cells and $135^3$ dark matter particles.
These simulations were run in an adiabatic fashion, i.e.,
no cooling/heating processes were included
except for shock heating, which is intrinsic to the simulations
and hence automatically included.
After the simulations were completed,
we calculated line emission and continuum emission
from heavy elements, in addition to those due to H and He,
simply by assuming that the metals should be present
in the cosmic gas, at the observed abundances.
The excluded dynamic effect of cooling/heating effect due to the metals,
which are not present in the original simulations,
is probably unimportant because
the cooling/heating time is typically much longer than
the dynamical time of the systems under consideration,
as we will show below.

The X-ray clusters in the simulations
are identified in exactly the same way
as before; see Kang \etal (1994) for details.
The metallicity in different (density) regions is assumed to have
the following form:
$$\eqalignno{Z(\rho_{sm}) &= 0.35\times
(1-e^{-(\rho_{sm}/\bar\rho_b)^2})\qquad, &(\the\eqtno) \cr}$$
where $Z(\rho_{sm})$ is in units of solar metallicity;
$\rho_{sm}$ is the baryonic density of the region in question
smoothed over a Gaussian window of radius $1h^{-1}$Mpc,
and
$\bar\rho_b$ is the global mean of the baryonic density.
This approximate formula is not intended to be very accurate,
but it suffices to our needs.
Equation (1) will produce for high density regions
the  metallicity observed in the great clusters, 0.35
(Edge \& Stewart 1991;
Arnaud \etal 1992,1994).
Then for regions with a low density, $\rho_{sm}$,
the metallicity declines rapidly, giving for gas in the voids
($\rho_{sm}/\bar\rho_b\approx 0.1$)
a metallicity of $3.5\times 10^{-2}Z_\odot$,
consistent with estimates/bounds obtained from studies of QSO
absorption lines (Fan \& Tytler 1994; Cowie \etal 1995).

The X-ray emissivity is calculated with a code based on the work
of Raymond \& Smith (1977), which
was kindly made available to us by
J.C.Raymond.  It is assumed that the gas is in
ionization equilibrium and optically thin.  The photoionization
by the diffuse radiation is ignored for the present study,
since collisional ionization
dominates over photoionization for $T>>10^5K$.
The solar abundances are taken from the CLOUDY code (Ferland 1994):
C (8.56); N (8.05) ; O (8.93); Ne (8.09); Mg (7.58); Si (7.55);
S (7.21); Ar (6.56); Ca (6.36); Fe (7.67); Ni (6.25).
The He abundance is assumed to be 10.90,
consistent with standard light-element nucleosynthesis.
The emission rates from H, He and metals are calculated separately.
Then each component is weighted by the relative abundance
and the sum of all components is our total emissivity.
We use the following energy bins to generate the
spectra:
0.5 eV for $h \nu=1-100$ eV, 5 eV for $h\nu=100-1000$ eV, 20 eV for
$h\nu= 1-10$ KeV, 40 eV for $h\nu=10-50$ KeV, and 100 eV for
$h\nu=50-1000$ KeV.

\medskip
\centerline{3. RESULTS AND COMPARISONS WITH OBSERVATIONS}
\smallskip
\centerline{\it 3.1 Luminosity Functions}
\medskip
Figures 1 and 2 show the X-ray
luminosity function for clusters of galaxies
in the two models considered: SCDM and LCDM.
Figures 1a-d, for the SCDM model
break the total luminosity into standard spectral bands
and compare our previous results
based on bremsstrahlung emission from a (H, He) plasma
(solid squares) to the full (line and continuum emission
(open circles) from a CLOUDY (Ferland 1994) mixture of elements.
Two points are to be noted.
First, for high luminosity
clusters $(L_x\ge 10^{44}$erg/s)
the differences are minimal, becoming significant
($\ge$ factor of three) only
at the lowest luminosities ($L_x\approx 10^{41}$erg/s),
where our observational knowledge is poor.
This is simply a manifestation of the temperature luminosity
relation noted in both observations
(Edge \etal 1990;
Henry \& Arnaud 1991; Henry 1992)
and in the numerical simulations (Cen \& Ostriker 1994).
At lower luminosities the clusters have lower temperatures,
leaving the atoms less fully ionized
and more capable of line emission.
Second, the differences are least in the hard bands
such as the 1-10keV range
(Figure 1d)
and most in the bands shown
in Figures 1b and 1c which reach
below 1keV, where the principal line emission occurs.

Figures 2a-2d show the same simulated
data from the LCDM model and show the same differences between
continumm only and line plus continuum computations.
The increase
due to line emission is greater for LCDM than
for SCDM, since the LCDM clusters are cooler.
Overall, out prior conclusions that SCDM over-predicts the number
of bright X-ray clusters but LCDM is in rough agreement
with observations remains true, as the high luminosity
clusters, where comparisons to observations are cleanest,
are not much effected by line emission.

\medskip
\centerline{\it 3.2 Spectral Energy Distributions}
\medskip
Figures 3 and 4 show detailed spectra
for the two models broken up in a different way.
All the luminosity emitted in the box from
regions within a $1h^{-1}$Mpc of the center
of a rich cluster having $L_x(bolometric)>10^{43}$erg/s
is shown (the solid line) as the mean spatial emissivity
due to ``X-ray clusters" in erg/cm$^3$/se/eV.
But much of the hot, emitting gas is further from cluster cores,
or in poor clusters or groups, or along
filaments and sheets.
These regions typically have such a low surface brightness
that
current and planned instrumentation does not detect
them as distinct ``sources".
But they should be seen in all sky studies and will have
specific spectral and angular distribution features.
The gas causing this residual emission is, of course,
highly clumped but it fills a much larger fraction of the total
volume than the $(1.8\times 10^{-4},6.4\times 10^{-5})$
filling factor occupied by the luminous clusters, in (SCDM,LCDM) models
respectively.
This gas is also, of course,
typically at a lower temperature than the several keV
gas in the great clusters.
In Figures 3a and 3b the dashed lines show
the volume-averaged
emissivity from this ``background" gas.
The upper panels (3a,3b)
show the spectra at a resolution which roughly corresponds
to the ASCA experiment (Tanaka, Inoue, \& Holt 1994)
with the relative displacement of curves
shown as computed.
The lower panels show the same information with arbitrary
shift of the vertical scale and higher
resolution on the horizontal scale to allow
better identification of individual lines
and to indicate what future experiments may find.

We note that in both models the background hot gas dominates
over the emission from rich clusters below
approximately 1keV.
Since our problems of spatial
resolution are more severe for the smaller structures producing
the background than
they are for computing the properties of the great
clusters, we expect that an accurate
calculation  might
increase the cluster emissivity by perhaps a factor of two and
the background by at least a factor of four, leading
to still greater dominance
of the background gas in the range
10eV$<h\nu< 10^3$eV.
The ratio is larger for the LCDM model than for
the SCDM model, because the former
has relatively lower temperatures.

The steepening of the spectrum below 1keV
seen recently by ASCA (\cf Gendreau \etal 1994)
as well as the OVII lines (also noted by ASCA)
discussed below are tentative but direct observational evidence
that the background gas we are discussing has
already been detected.

This new component of soft X-ray emitting gas is only apparent
when we allow properly for line emission.
For example, if in Figure 4a,
we artifically assume
zero metallicity for the background gas we reduce
the luminosity in the 10-1000eV
by a factor of $3.1$.
The spectral signature of the background as seen
in the low resolution spectra,
Figures 3a and  4a
is the  ``iron bump" (actually a mixture of iron and oxygen)
in the region 0.5-1.0keV.

Now let us turn to
Figures 3b and 4b which show detailed spectra that allow us to identify
better those spectral
lines which are contributing most to the background.
We see
in the background gas a strong signature in the 0.561-0.574keV
(marked with ``*") range from a blend of OVII lines
(characteristic of $10^6$~K gas)
that is weak or absent
in the clusters emission spectra.
Also strong
in the background gas is the 0.904-0.922keV
blend of lines from NeIX at a similar
temperature.
Both features are nearly absent
in the cluster gas.
The strong OVIII line at 0.654keV (marked with ``o")
seen in both cluster and background gas
will give a separate indication of oxygen
abundance allowing the ratio
of the 0.57keV to the 0.65keV features
to be used as a temperature indicator.
We have found one other spectral feature,
which is prominant in the background gas,
but is absent from the cluster spectra.
This is a pair of iron XVII lines
at 0.726keV and 0.739keV (marked with ``x")
which have peak strength in gas at approximately
$10^{6.5}$K.
These are strong in the lower (background) dashed curves
of Figures 3b and 4b, but are not apparent
in the cluster gas, where all the strong Fe lines
are from higher ionization species, typically Fe XIX$\rightarrow$ Fe XXII.

Figures 5a,b show typical spectra
(from the LCDM model)
for a high temperature (2.7keV) high luminosity ($1.6\times 10^{44}$erg/s)
cluster and a lower temperature (1.8keV) lower luminosity
($3.5\times 10^{43}$erg/s) cluster.
Overall, line emission is stronger in the lower temperature cluster
particularly the
$\sim 1$keV iron blend and the 0.654keV OVIII line.

How can
the background gas that we are discussing be distinguished
from the hot Galactic gas along the same line of sight?
To some extent the soft X-ray
``background" that we are proposing is the same as Galactic emission
in that the sum over all galaxies like our own
does make a nontrivial
contribution to the soft X-ray background.
If the average galaxy emits $2\times 10^{39}$erg/s in
the 0.5-1.5keV band (\cf Fabbiano 1989),
then the sum of all such will
produce an emissivity of $10^{-39}$erg/cm$^3$/sec/eV
not far from the levels shown in Figure 4a.
But most of the emission shown by the dashed line in 4a
is due to hot gas in bigger systems than our own galactic disk or halo.
Thus, the temperature is in the range $10^6-10^7$~K,
whereas, for most of the galactic coronal gas,
the typical temperature
(as weighted by $\rho^2$)
is in the range $10^5-10^6$~K.
Thus, the line ratios indicative of the
OVI/OVII and OVII/OVIII ratios will help in distinguishing
between the two components.
The mean logarithmic slope of
$S_\nu$ will also provide an important discriminant,
in that the background emission shows a much
steeper slope than the cluster emission below 1keV.

But both  the line ratios
and the slope of the continuum will be effected by the convolution
over redshift, i.e., we see along any line of sight radiation
emitted at a variety of epochs shifted by various amounts.
However, this effect, which will blur out line
features, is not as large
as might be expected.
Examination of Figure 14 of the LCDM paper shows
that in even the softer (0.3-3.5keV)
band half of intensity arises from redshift of less
$z=0.32$.
Thus, the mean redshift of the emitted $\sim$1keV
photons will be close to $z=0.32$
with a dispersion of perhaps $\sigma=0.32$.
As a consequence,
the blurring of spectral features in the background will be approximately
0.2/1.3=24\% with a shift of 32\%.
Specifically the iron blend should move from the region 0.6-1.0keV
(FWHM) to the region 0.45-0.81 keV with a width only slightly
larger than shown in Figure 4a.

In fact, a strong soft X-ray background radiation field
has been detected
(see McCammon \& Sanders 1990)
but, for a variety of reasons, the very low energy
part of this is believed to be of Galactic rather than extragalactic origin.
For example, in the C-band, 0.2-0.3 keV,
the minimal shadowing
of the soft X-ray background by SMC
(recently reviewed by McCammon \& Sanders 1990)
seriously limits the extragalactic component,
showing that a large fraction originates from a hot
component of the ISM (\cf Mckee and Ostriker 1977),
or a hot galactic halo.
But there are reasons to believe that a significant
fraction of at least the component in the range 0.5-1.0keV
is truly extragalactic (\cf Burrow \& Kraft 1993;
Gendreau \etal 1994).
The spectral features which we noted
earlier should help provide
clues to the origin of the background radiation in this range
as the galactic component is probably
too cool to produce
the iron blend which should be prominent in the
background component
described in this paper.

\medskip
\centerline{\it 3.3 Cooling Cluster Cores}
\medskip
Now let us turn to the question of cooling flows.
There is abundant eivdence (Fabian 1994) that
in the central regions of many clusters substantial
amounts of gas exist that have cooling times which are short
compared to the Hubble time.
We have not addressed this issue in the past
for two reasons:
first, since we did not include metal line emission,
our cooling functions were inaccurate in just those regions
where cooling might be most  important;
and, second, our spatial resolution was inadequate
to examine accurately the dense central regions which are
most likely to cool.
We can address the first of these issues now, but our
resolution, while improved over our earlier calculations,
is still not sufficient to address the
problem accurately so our remarks must be confined
to comparative issues.
For which models and which luminosity ranges is cooling
of greatest importance?
Figures (6) and (7) show the ratio of the Hubble
time to the cooling time for the central cell in each cluster,
the first figure showing the distributions of this ratio for the SCDM model
and the second the distributions for the LCDM model.
The volume of the central cell is $(315k^{-1}kpc)^3$ which
corresponds to a sphere of radius $195h^{-1}$kpc.
Given the unavoidable numerical diffusion,
a more accurate
estimate of our effective central cell radius might be approximately
$250h^{-1}$kpc.
In both Figures (6) and (7)
we divide the clusters into three groups based on the total
bolometric luminosity.
We see that for the highest luminosity clusters ($L_{bol}>10^{43}$erg/s)
the computed
cooling times are very long on average compared to the Hubble time.
For the SCDM model $t_H/t_{cool}$ is always less than $0.03$
and for the LCDM model it is less than $0.06$.
However our computed sample size is small and it seems likely that the
tail would reach to a value of $t_H/t_{cool}$ which is
factor of 10 higher were the sample size to include 100 bright
clusters.
This issue of a small statistical sample plus our relatively poor
resolution make it likely that perhaps a few percent of
the bright clusters would have cooling cores
in perfect simulations of these two models.
Work currently in progress should greatly clarify this issue.

But we can say with some certainty that the lower luminosity
clusters, since they have lower temperatures,
cool much more efficiently.
This is especially evident in the LCDM model (Figure 7c),
where it is clear that cooling would be an
important effect for higher resolution simulations.

The trend for the increased importance
of cooling in Figures (6) and (7) as we go from
(a) $\rightarrow$ (c) is evident
(low luminosity regions are more subject to cooling than high luminosity
regions)
and strengthens
the conclusions of section (3.1).
When cosmic gas cools, it typically contrasts to a higher
density and thus cools still more rapidly.
Thus, if we had allowed for cooling in the evolution of the gas,
then the background emission that we found would have
been still stronger with respect to the average cluster emission.

\medskip
\centerline{4. CONCLUSIONS}
\medskip

Straightforward physical theory predicts that the gravitationally
induced growth of structure will lead to an accumulation
of hot shocked gas at the vertices
where caustics intersect.
Numerical simulations have in fact confirmed that some
models for the growth of cosmological structure do fit
rather nicely the X-ray observations
of clusters of galaxies.
However the same theories also predict the
existence of much more
hot gas, at a somewhat lower temperature, filling a significant
fraction of cosmic space with X-ray  emitting
sheets and filaments.
In this paper we compute the spectral properties of this
``background" gas noting that it
should dominate over the cluster gas (in total emissivity)
at energies below $1$keV.

Figures (8) for SCDM and (9) for LCDM summarize
the nature of the regions emitting the two components
at redshift zero,
the upper panels showing the volume weighted distribution
of $\rho^2$ (as a function of gas temperature)
and the lower panels the emissivity-weighted
($\int^{10{\rm keV}}_{200{\rm eV}} S_\nu d\nu$)
distribution.
Examining Panel (9b) we see that the cluster X-rays
are typically emitted from gas
in the range $10^{7.0-7.8}$K whereas
the background gas emission is broadly distributed in the range
$10^{6.0-7.3}$K.
The signature of the background gas will be softer
spectrum,
certain specific lines and blends, and a spatial extent
which is much more diffuse (in terms of angular
autocorrelation function)
than the cluster gas.
The long standing problem that the spectrum of the background
does not match well to the spectrum of the resolved sources
remains with us.
Further as pointed our by Hasinger \etal (1993) and
earlier by Hamilton and Helfand (1987)
there is room, even perhaps a necessity for either a
very faint unclusterd population of sources
or a more smoothly distributed component.
Also recent ASCA measurements (Gendreau \etal 1994)
indicate the existence of a background component with a
steep spectrum below 1keV and evidence from OVII lines.
Thus, it may be that ROSAT, ASCA and other satellites have already observed
the emission from this background gas but,
because of its diffuse nature,
rather specific analyses of the observational
data must be undertaken to confirm the detection.

We are happy to acknowledge
support from grants NAGW-2448, NAG5-2759, AST91-08103
and ASC93-18185.
It is a pleasure to acknowledge NCSA for allowing
us to use their Convex-3880 supercomputer.
We are extremely grateful
to John C. Raymond who
generously provided for us the computer
code that enables us to compute line and continuum
emission rates for heavy elements.
Discussions with
H. Bohringer,
B. Draine, G. Hasinger, R. Mushotzky,
B. Paczynski,
A. Soltan and J. Trumper,
are gratefully acknowledged.

\vfill\eject

\centerline{REFERENCES}
\refset
\smallskip
Arnaud, M., Rothenflug, R., Boulade, O., Vigroux, L., \& Vangioni-Flam, E.
1992,
A\& A, 254, 49
\refset
\smallskip
Arnaud, K., Mushotzky, R.F., Ezawa, H., Fukazawa, Y., \etal 1994, preprint
\refset
\smallskip
Bryan, G.L., Cen, R., Norman, M.L., Ostriker, J.P., \& Stone, J.M. 1994, ApJ,
428, 405
\refset
\smallskip
Burrows, D.N., \& Kraft, R.P. 1993, ApJ, 411, 685
\refset
\smallskip
Cen, R., \& Ostriker, J.P. 1994, ApJ, 429, 4, LCDM
\refset
\smallskip
Cen, R., Ostriker, J.P., Soltan, A.M., \& Hasinger, G. 1995, preprint
\refset
\smallskip
Cowie, L.L., Songaila, A., Kim, T.S., \& Hu, E. 1995, AJ, in press
\refset
\smallskip
Edge, A.C, Stewart, G.C., Fabian, A.C., \& Arnaud, K.A. 1990, MNRAS, 245, 559
\refset
\smallskip
Edge, A.C., \& Stewart, G.C. 1991, MNRAS, 252, 428
\refset
\smallskip
Fabbiano, G 1989, ARAA, 27, 89
\refset
\smallskip
Fabian, A.C. 1994, ARAA, 32, 277
\refset
\smallskip
Fan, X.-M., \& Tytler, D. 1994, ApJS, 97, 17
\refset
\smallskip
Ferland, G.J. 1994, preprint
\refset
\smallskip
Gendreau, K.C., Mushotzky, R., Fabian, A.C.,
Kii, T., Holt, S.S., Serlemitsos, P.J.,
Tanaka, Y., Ogasaka, Y., Bautz, M.W.,
Fukazawa, Y., Ishisaki, Y., Kohmura, Y.,
Kakishima, K., Tashiro, M., Tsusaka, Y.,
Kunieda, H., Ricker, G.R., \& Vanderspek, R.K. 1994, preprint
\refset
\smallskip
Hasinger, G., Burg, R., Giacconi, R., Hartner, G., Schmidt, M., Trumper, J.,
\& Zamorani, G. 1993, Astron. Astrophys., 275, 1
\refset
\smallskip
Hamilton, T.T., \& Helfand, D.J. 1987, ApJ, 318, 93
\refset
\smallskip
Henry, J.P. 1992, in ``Clusters and Superclusters of Galaxies",
 d. A.C. Fabian, 311, (Kluwer Academic Publisher).
\smallskip
\refset
Henry, J.P., \& Arnaud, K.A. 1991, ApJ, 372, 410
\smallskip
\refset
Kang, H., Cen, R., Ostriker, J.P., \& Ryu, D. 1994, ApJ,
428, 1
\smallskip
\refset
McCammon, D., \& Sanders, W.T. 1990, ARAA, 28, 657
\smallskip
\refset
Mckee, C.F., \& Ostriker, J.P. 1977, ApJ, 218, 148
\smallskip
\refset
Raymond, J.C., \& Smith, B.W. 1977, ApJS, 35, 419
428, 1
\smallskip
\refset
Tanaka, Y., Inoue, H., \& Holt, S.S. 1994, PASJ, 46, L37
\smallskip
\refset
\vfill\eject

\centerline{FIGURE CAPTIONS}
\medskip

\item{Fig. 1--}
Figures (1a,b,c,d) show
the computed luminosity functions in four different bands:
bolometric, $0.3-3.5$keV,
$0.5-4.5$keV,
$2-10$keV, respectively,
for the SCDM model at $z=0$.
The open symbols include line emissionsas well as bremsstrahlung;
the solid squares include only bremsstrahlung due to H and He.
The hatched region in Figure (1a) indicates observations
by Henry and Arnaud (1991) and that in (1d) the observations
of Henry (1992).

\item{Fig. 2--}
Analagous to Figure 1 but for the LCDM model.

\item{Fig. 3--}
Panel (a) shows volume-averaged
X-ray emissivity of all the bright clusters $L_{bol}>10^{43}$erg/s
(solid curve) and X-ray emissivity of the
background (dashed cruve) at redshift $z=0$ for the SCDM model.
We use the spectral resolution of ASCA SIS detector, which
has resolution of 60~eV at 1~keV and 140~eV at 8~keV (linear
interpolation is used for the range 0.4~keV to 9~keV which SIS
detector is sensitive, below 0.4keV we simply assume a spectral
resolution of 60eV).
Panel (b) shows a high resolution version of panel (a)
with the two curves displaced vertically by an arbitrary amount
for display purpose.
Note the prominant ``iron bump" in the background emission.
In panels 3b and 4b
of the high resolution spectrum we designate three
temperature sensitive lines
by ``*" (from an OVII
blend characteristic of $10^6$K gas at $0.561-0.574$ keV),
by ``o" (from OVIII of $10^{6.5}$ K gas at 0.654 keV)
and
by ``x" (a pair of iron XVII lines at 0.726 keV and 0.739 keV from
gas at $10^{6.5}$K).

\item{Fig. 4--}
Analagous to Figure 3 but for the LCDM model.
Note how the background emissivity dominates
over the radiation from clusters below 1keV.

\item{Fig. 5--}
Figures (5a,b) show two spectra of two bright X-ray clusters in the LCDM
model.

\item{Fig. 6--}
Figures (6a,b,c) show histograms of
the ratio $t_{Hubb}/t_{cool}$ for X-ray clusters
in three different luminosity bands, in the SCDM model.
model.

\item{Fig. 7--}
Analagous to Figure 6 but for the LCDM model.
Cooling should be important in low luminosity clusters.

\item{Fig. 8--}
Figure 8 shows a plot of $\rho^2$ on the cell scale
as a function of temperature
in two different ways:
volume-weighted (top panel) and
weighted by emissivity
($\int^{10{\rm keV}}_{200{\rm eV}} S_\nu d\nu$) (bottom panel),
for the SCDM model at $z=0$.
Each panel is further broken
into two different components: clusters and background.
Note that the background radiation is emitted by considerably
lower temperature gas than that
producing the cluster emission.

\item{Fig. 9--}
Figure 9 is the same as Figure 8 but for the LCDM model.

\vfill\eject
\bye